\documentclass[twocolumn,superscriptaddress,showpacs,preprintnumbers,prl]{revtex4}

\usepackage{graphicx}

\newcommand{\twid}{\sim}

\newcommand{\gap}{\hspace{.4in}}

\newcommand{\gt}{\rightarrow}

\newcommand{\period}{\ \ .}
\newcommand{\comma}{\ ,\ }

\newcommand{\lsim}{\,\stackrel{<}{\scriptstyle \sim}\,}

\newcommand{\ignore}[1]{}

{\begin{tabular}{#1}}%
{\end{tabular}}

{\end{list}}

{\end{list}}

{\ \\ XXXXX \hfill XXXXX #1 XXXXX\begin{equation}\label{#1}}%
{\end{equation}}

{\ \\ XXXXX \hfill XXXXX #1 XXXXX\begin{eqnarray}\label{#1}}%
{\end{eqnarray}}

\newenvironment{bequation}[1]%
{\begin{equation}\label{#1}}%
{\end{equation}}

\newenvironment{beqnarray}[1]%
{\begin{eqnarray}\label{#1}}%
{\end{eqnarray}}

\newcommand{\drop}{\nonumber \\}

\newcommand{\ie}{i.\,e.~}

{\begin{center} \section*{#1} \end{center}}

\newcommand{\degreesc}{$^0$C }

{\end{figure}} 

{\end{figure}}

{\end{figure}}

\newenvironment{eq}[1]%
{\begin{bequation}{#1}}{\end{bequation}}

\newenvironment{eqarray}[1]%
{\begin{beqnarray}{#1}}{\end{beqnarray}}

\def\eqref#1{(\ref{#1})}

\newcommand{\Nbar}{\bar{N}}
\newcommand{\Nbart}{\bar{N}_t}

\newcommand{\phit}{\phi_t}
\newcommand{\vt}{v_t}
\newcommand{\Dt}{D_t}
\newcommand{\mt}{m_t}

\newcommand{\vplus}{v^+}
\newcommand{\kplus}{k^+}
\newcommand{\vminus}{v^-}

\newcommand{\tauqs}{\tau_{\rm qs}}

\newcommand{\taufill}{\tau_{\rm fill}}
\newcommand{\tstar}{t^{*}}

\newcommand{\meq}{m_{eq}}
\newcommand{\phieq}{\phi_{eq}}

\newcommand{\rhochains}{\rho_{\rm chains}}
\newcommand{\Deq}{D_{\rm eq}}
\newcommand{\veq}{v_{\rm eq}}

\newcommand{\mtotal}{m_{\rm total}}
\newcommand{\Nbareq}{\bar{N}_{eq}}
\newcommand{\vtdot}{\dot{v}_t}

\newcommand{\phidot}{\dot{\phi}}

\newcommand{\Tp}{T_p}

\newcommand{\taufast}{\tau_{\rm fast}}
\newcommand{\tauslow}{\tau_{\rm slow}}

\newcommand{\cubed}{$^3$}

\begin{document}

\preprint{submitted to {\em Phys. Rev. Lett.}}

\title{The Ultrasensitivity of Living Polymers}

\author{Ben O'Shaughnessy}

\affiliation{Department of Chemical Engineering, Columbia University, New York, NY 10027}

\author{Dimitrios Vavylonis$^{1,}$}

\affiliation{Department of Physics, Columbia University, New York, NY 10027}


\begin{abstract}

Synthetic and biological living polymers are self-assembling chains
whose chain length distributions (CLDs) are dynamic. We show these
dynamics are ultrasensitive: even a small perturbation
(e.g. temperature jump) non-linearly distorts the CLD, eliminating or
massively augmenting short chains. The origin is fast relaxation of
mass variables (mean chain length, monomer concentration) which
perturbs CLD shape variables before these can relax via slow chain
growth rate fluctuations. Viscosity relaxation predictions agree with
experiments on the best-studied synthetic system,
$\alpha$-methylstyrene.

\end{abstract}



\maketitle


The term ``Living Polymers'' labels a remarkable and diverse family of
self-assembling systems in which molecules or other microscopic units
spontaneously aggregate into long chains by continuously adding to
chain ends. Their technological and biological importance and their
unusual properties as examples of soft condensed matter have driven a
large body of experimental and theoretical
research\cite{greer:review,webster:szwarc:prl_combo,%
schafer:living_theory_prl_combo,miyakestockmayer:living,%
das:greer:living_7_mwd_etal,zhuang:greer_kinetics_etal,garcia:greer:living_nonumber%
,patten:livingfrp_prl_short,pollard:living_biorefs_prl_combo,%
frischknechtmilner:living_aggregation,aniansson:micelle_kinetics_theory_expt_etal,
livingpol_wlike_dynamics}. The classic synthetic example is ionic
living polymerization\cite{greer:review,webster:szwarc:prl_combo,%
schafer:living_theory_prl_combo,miyakestockmayer:living,%
das:greer:living_7_mwd_etal,zhuang:greer_kinetics_etal,garcia:greer:living_nonumber}
where monomers assemble into flexible polymer chains by adding to
charged chain ends which remain ``alive'' even after monomer is
consumed, a property widely exploited commercially to synthesize high
performance block copolymers and other novel
materials\cite{webster:szwarc:prl_combo}.  A recent variant on this
theme with immense technological potential is living free radical
polymerization\cite{patten:livingfrp_prl_short} where ingenious
capping-decapping schemes prevent termination and yet permit chain
propagation. In the biological world, actin and microtubule filaments,
rapidly assembled from the proteins actin and tubulin, are essential
to the motility and structural integrity of living cells
\cite{pollard:living_biorefs_prl_combo}.

As polymeric or filamentous materials, the novel and distinguishing
feature of these systems is that unlike inert polymers the chains are
{\em dynamic} or ``living'' objects whose lengths constantly
fluctuate.  The polymerization processes are alive: when external
conditions change, chain length distributions can respond dynamically
and relax to a new equilibrium.  This adaptability is the crucial
property exploited in both synthetic and natural applications. For
example, it enables living cells to rapidly initiate motion or shape
changes by altering cellular conditions in response to extracellular
signals.

In this letter we study this dynamical responsiveness theoretically.
Whilst a rather clear picture of equilibrium properties
\cite{greer:review,schafer:living_theory_prl_combo} has been
established, living polymer dynamics are far less well understood
\cite{miyakestockmayer:living,das:greer:living_7_mwd_etal,zhuang:greer_kinetics_etal,garcia:greer:living_nonumber}. We
focus on the synthetic system which has received most experimental
attention\cite{greer:review,das:greer:living_7_mwd_etal,zhuang:greer_kinetics_etal,garcia:greer:living_nonumber},
the anionic living polymer poly $\alpha$-methylstyrene (PAMS) and we
compare our predictions to PAMS viscosity relaxation measurements at
the end of this letter.  Aside from its intrinsic importance, PAMS is
a model system for more complex cases such as biological living
polymers and most of our results are completely general.  We use the
term ``Living Polymers'' in the traditional sense
\cite{greer:review,webster:szwarc:prl_combo} to denote systems where
(i) the concentration of living chains, $\rhochains$, is fixed for all
time by the number of initiators, and (ii) chains grow at their ends
only.  Related classes include systems where surfactants aggregate
into spherical \cite{aniansson:micelle_kinetics_theory_expt_etal} or
elongated (``wormlike'') micelles \cite{livingpol_wlike_dynamics}
whose number is not fixed, and whose
dynamics\cite{aniansson:micelle_kinetics_theory_expt_etal,livingpol_wlike_dynamics}
are very different.

\ignore{
and we conclude with a brief discussion on their
broad relevance, including how living cells may harnass the effects we
identify to their advantage.
} 

\begin{figure}
\includegraphics[width=7.5cm]{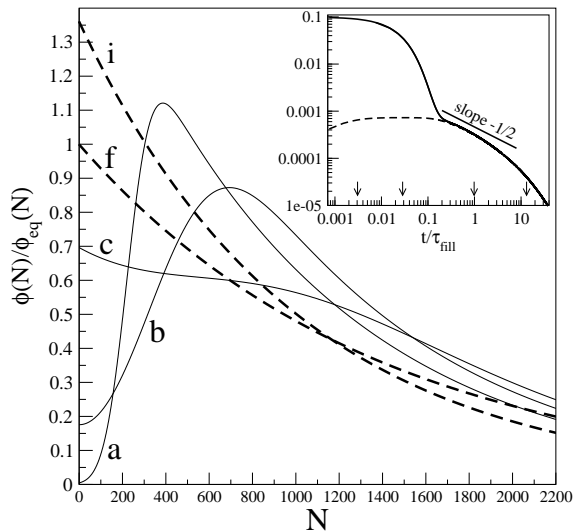}
\caption{\label{mwd} Numerical results for response of CLD to a small
$T$-jump perturbation of amplitude $\epsilon =0.1$ perturbing
T-dependent parameters $\vminus$ and $\kplus$ in eq. \eqref{cc}.
Final equilibrium exponential (curve f, $\Nbareq = 1366, \theta =
0.37$) is close to initial equilibrium exponential (curve i).  Along
its path $i\gt f$ the CLD deviates non-linearly from both.  Times in
units of $\vminus$ (for PAMS \cite{greer:review}, $\vminus \approx 0.1
s$). Timescales: $\tstar \approx 400, \taufast \approx 3640, \taufill
\approx 1.32 \times 10^{5}, \tauslow \approx 1.86 \times
10^{6}$. Curve a, $t = 3640$: a hole has appeared at small $N$.  Curve
b, $t = 3 \times 10^4$: hole-filling. Curve c, $t = 2.3 \times 10^5$:
final shape relaxation towards equilibrium, curve f.  Inset:
Relaxation of fast and slow variables. Arrows from left to right
indicate $\tstar, \taufast, \taufill, \tauslow$.  Fast variables
$\delta \mt / \meq = -\theta\, \delta \Nbart/\Nbareq$ (solid)
decay exponentially after $\taufast$, and are then enslaved to the
slow variable $-\delta \phit(0)$ (dashed).  Note initial
nonlinear increase of $- \delta \phit(0)$ after $\tstar$.  Theory
predicts $\delta \phit(0) \twid t^{-1/2}$ during $\taufill < t <
\tauslow$ and $\delta \phit(0) \twid e^{-t/\tauslow}$ for $t >
\tauslow$.}  \end{figure}

The principal conclusion of this work is that living polymers are {\em
ultrasensitive}, \ie highly dynamically susceptible to small
perturbations. Thus a small change in external conditions, inducing a
small change in the equilibrium state well-described by a linear
susceptibility, has nonetheless a large dynamical effect: intermediate
states deviate strongly from equilibrium in that some observables are
perturbed in a highly non-linear manner.  To quantify this, consider
PAMS whose equilibrium chain length distribution (CLD) is close to the
broad exponential
\cite{greer:review,das:greer:living_7_mwd_etal} predicted
by theory, $\phieq(N) \twid e^{-N/\Nbareq}$ (see fig. \ref{mwd}), with
typical mean number of monomer units per chain $\Nbareq$ in the range
\cite{garcia:greer:living_nonumber} of a few 100 to several 1000.
A standard experimental procedure is the ``T-jump'' where temperature
is suddenly changed by an amount $\Delta T$.  Consider a PAMS system
with $\Nbareq =1000$ subjected to a small decrease $\Delta T=
-5$\degreesc as measured by the small parameter $\epsilon \equiv
-\Delta T/T_0 \approx 0.1$, where $T_0\equiv \partial T/\partial \ln
\meq \approx 50$\degreesc for typical PAMS studies and $\meq$ is the
equilibrium monomer concentration.  The theory presented here predicts
the onset after $\twid 3$ hours of a drastic depletion of short
chains. By 10 hours a hole has appeared in the CLD in the region
$0<N\lsim 400$ (see fig. \ref{mwd}) and lengths $N\lsim 100$ have
virtually disappeared. This is despite the fact that the change in the
equilibrium CLD $\phieq(N)$, which is recovered after $\twid 10^3$
hours, is destined to be small, $\delta \phieq(N) /\phieq(N) = {\cal
O}(\epsilon)$ for all $N$.  This hole completely invalidates
perturbation theory: {\em there are no small perturbations} (other
than those so tiny as to be beyond typical experimental resolution).
This is the essence of ultrasensitivity.

\begin{figure}
\includegraphics[width=6cm]{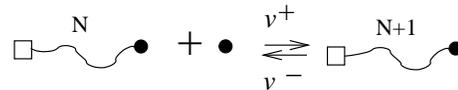}
\caption{\label{scheme} A living polymer is an initiator (square) plus
$N$ monomers.  Monomers can add (subtract) from its live end (filled
circle) at rates $\vplus$ ($\vminus$). The initiators of
refs. \onlinecite{garcia:greer:living_nonumber,das:greer:living_7_mwd_etal}
are bifunctional (one interior initiator plus two live ends).
}
\end{figure}

We consider living polymers, such as the PAMS systems studied by Greer
and coworkers
\cite{greer:review,das:greer:living_7_mwd_etal,zhuang:greer_kinetics_etal,garcia:greer:living_nonumber},
where monomers spontaneously polymerize below a ceiling temperature
$\Tp$, see fig. \ref{scheme}.  The dynamics of $\phi_{t,N}$, the
number of chains of length $N$, are $\phidot_{t,N} = \vplus
\phi_{t,N-1} - (\vplus + \vminus) \phi_{t,N} + \vminus \phi_{t,N+1}$
where $\vplus = \kplus m$ and $\vminus$ are monomer addition and
dissociation rates from chain
ends\cite{living:kplusvminus_independentofN_note} ($\kplus$ is the
addition rate constant). Rearranging terms and taking the continuous
limit immediately gives
                                                \begin{eqarray}{cc}
&& \partial \phit / \partial t = 
- \vt \partial \phit / \partial N + \Dt \partial^2 \phit /
\partial N^2
\comma \drop
&& \vt \phit(0)-\Dt [\partial \phit/\partial N]_{N=0} = 0 
\comma
                                                                \end{eqarray}
where a zero current boundary condition applies at $N=0$.  Note a
second derivative term naturally emerges, with a ``diffusivity''
coefficient $\Dt \equiv (\vplus + \vminus)/2$, representing
fluctuations about the average growth rate or ``velocity'' $\vt \equiv
\vplus - \vminus$.  Both $\vt$ and $\Dt$ depend on $\phit$ (see
eq. \eqref{milk} and following remarks).  Setting $\partial \phit /
\partial t = 0 $ gives the equilibrium exponential CLD with $\Nbareq =
- \Deq / \veq$.  Since $\Nbareq\gg 1$, it follows that in equilibrium
$\vplus \approx \vminus$, the diffusivity is $\Deq \approx \vminus$
and the velocity has a small negative value relative to the
characteristic scale $\vminus$
                                                \begin{eq}{bottle}
\veq = - \vminus / \Nbareq \period
                                                                \end{eq}
This is just sufficient to negate diffusive broadening which would
otherwise smear out the equilibrium MWD of width $\Nbareq$ after time
$\tauslow = \Nbareq^2/\Deq$; \ie, $\veq \tauslow = -\Nbareq$.  Note
that since the average velocity over all chains must vanish in
equilibrium (there can be no net dissociation to the monomer pool) the
$N>0$ chains must have a small negative velocity to balance the unique
$N=0$ chains (consisting of an initiator only, fig. \ref{scheme})
which cannot depolymerize and so have a {\em positive} velocity.

A clue as to the origin of ultrasensitivity is already apparent in
this equilibrium situation.  The significance of eq. \eqref{bottle} is
that equilibrium is an extremely delicate balance sustained by a tiny
negative velocity easily overwhelmed by even a very small
perturbation.  Consider an equilibrium system suddenly subjected to a
small negative T-jump $\Delta T$, after which the current monomer
concentration $m$ exceeds its equilibrium value for the new
temperature by $\delta m \approx -(\Delta T /T_0) \meq \equiv \epsilon
\meq$ ($\delta X$ denotes the deviation of any property $X$ from its
equilibrium value, $X_{\rm eq}$, which will eventually be attained
after relaxation).  This generates a velocity boost $\delta v = \kplus
\delta m$. Since $\kplus \meq \approx \vminus$ to order $1/\Nbareq$,
we have
                     \begin{eq}{camel}
\delta v \approx \epsilon \vminus \period
                                                                \end{eq}
Comparing eqs. \eqref{bottle} and \eqref{camel} one sees that despite
the smallness of the perturbation, $\epsilon\ll 1$, the delicate
balance of velocity and diffusion which characterizes equilibrium is
destroyed since the boost $\delta v$ greatly exceeds the equilibrium
velocity $\veq$. This holds for any $\epsilon$ above a tiny threshold
$1/\Nbareq$.  Immediately after the perturbation, diffusive broadening
$\twid (\Deq t)^{1/2}$ is still stronger than coherent chain growth
$\twid \epsilon \vminus t$. But for times larger than $\tstar \approx
\Deq/\epsilon^2\vminus$ chains grow coherently, \ie $\Nbar$ will
increase until the excess mass $\delta m$ held by the monomer
reservoir has been transferred to the polymer system and the velocity
boost has decayed.  In the process the mean chain length increases by
an amount $-\delta \Nbar$ determined by conservation of total
concentration of monomers $\mtotal$:
                       \begin{eqarray}{place}
&m& + \rhochains\, \Nbar = \mtotal \comma\ \ \  
      \delta \Nbar/ \Nbareq = -\epsilon\,\theta^{-1} \period
                                                                \end{eqarray}
The parameter $\theta\equiv (\mtotal-\meq)/\meq \approx (T_p-T)/T_0$
measures distance into the polymerization regime, and will be taken as
order unity here. Thus, the entire CLD translates uniformly by
$\approx \epsilon \Nbareq$ (fig. \ref{mwd}) in time
                     \begin{eq}{potato}
\taufast = \delta \Nbar/(\epsilon\vminus) = 
			 \Nbareq/(\theta \,\vminus)
							\period
                                                                \end{eq}
This leaves behind a hole in the CLD: chains with lengths less than
$\epsilon \Nbareq$ have disappeared.

For times beyond $\taufast$, mass transfer is essentially complete and
$m, v$ and $\Nbar$ are very nearly relaxed.  These are the fast
variables.  The process of CLD {\em shape} relaxation needs much more
time. It relies on the far slower diffusive process of incoherent
reshuffling of monomers between chains.  The time $\taufill$ needed to
fill the hole is simply the diffusion time corresponding to the hole
width,
                     \begin{eq}{hand}
\taufill = \epsilon^2 \tauslow/\theta^2 \period
                                                                \end{eq}
The last process is global CLD shape relaxation on the scale
$\Nbareq$, requiring a diffusion time $\tauslow=\Nbareq^2/\Deq$.
These events are depicted in fig. \ref{mwd}.

In summary, relaxation to the new equilibrium state involves 3
distinct episodes: (i) coherent chain growth for $0<t<\taufast$ during
which fast mass variables relax, (ii) hole-filling, \ie recovery of
short chains, during $\taufast<t<\taufill$ and (iii) global diffusive
relaxation for $t>\taufill$ during which slow shape variables relax on
a timescale $\tauslow$.  It is this conflict of timescales which is
the origin of the non-linear hole produced by (i): slow diffusive
shape equilibration simply cannot keep pace with the rapid deformation
produced by mass transfer.

It is interesting to compare this with spherical micelle aggregation
which involves two distinct relaxation processes
\cite{aniansson:micelle_kinetics_theory_expt_etal}. The first entails fast
mass exchange between monomers and nearly monodisperse micelles whose
number remains essentially fixed due to large aggregate
nucleation/dissociation barriers. This is similar to process (i) above.
However, a crucial difference is that the number of aggregates can
ultimately change and the timescale for this process in consequence
depends inversely on total monomer concentration (whereas $\taufast$
is independent of $\mtotal$). During a second, much slower process,
the number of micelles re-equilibrates.  This process has no analogue
for the living polymers we study.

Let us now briefly outline formal calculations justifying these
arguments, starting from eq. \eqref{cc}. Its steady state solution is
the equilibrium Flory-Schultz \cite{greer:review} CLD, $\phieq(N) =
e^{-N/\Nbareq} /\Nbareq$ with $\Nbareq \approx
\mtotal\theta/[\rhochains(1+\theta)]$.  Multiplying eq. \eqref{cc} by
$N$ and integrating one has
                                                \begin{eq}{milk}
\vtdot = - \vt/\taufast -  \Dt \phit(0)/\taufast \comma
                                                                \end{eq}
after using eqs. \eqref{place}
and \eqref{potato}.  Given the CLD
$\phit$, this relationship determines the time-dependent velocity
$\vt$ and thence diffusivity $\Dt=\vt/2 + \vminus$.  The technical
difficulty is that both $\vt$ and $\Dt$ depend non-locally on $\phit$,
\ie eq. \eqref{cc} is a non-linear and non-local system.

Consider a negative T-jump perturbation inducing, as discussed, a
velocity $v_0 \approx \epsilon \vminus$ at $t=0$.  Now, to order
$\epsilon$, $\Dt \approx \vminus \approx \Deq$.  Thus in the velocity
kinetics, eq. \eqref{milk}, the $\vt$ term is initially much
greater than the $\Dt$ term (provided $\epsilon > 1/\Nbareq$). Hence
$\vt \approx \epsilon \vminus e^{-t/\taufast}$ in the CLD evolution
kinetics eq. \eqref{cc} in which, for $t>\tstar$, the coherent term
wins and the CLD translates along the N axis, $\phit(N) \approx
\phi_0(N-\int_0^t \vt)$. Its displacement converges for $t\gg
\taufast$ to $\epsilon \Nbareq/\theta$.  In other words the CLD
translates and then halts on a timescale $\taufast$, leaving a hole of
size $\approx \epsilon \Nbar$ in its wake (see fig. \ref{mwd}).  More
precisely, the trailing edge broadens by $~(\Deq t)^{1/2}$ and
produces an exponentially small amplitude at the origin, $\phit(0)
\twid e^{-t/\tstar}$.

This concludes episode (i). The linearly related fast variables $v, m$
and $\Nbar$ have the same decay kinetics and for $t\gg\taufast$ have
all relaxed. Since the velocity is exponentially small, the CLD
kinetics eq. \eqref{cc} now describe essentially pure diffusion with
reflecting boundary conditions.  This is episode (ii): on the
timescale $\taufill$, diffusion fills the hole (see fig. \ref{mwd})
and replenishes the amplitude at the origin, $\phit(0) \approx
(t/\taufill)^{1/2} e^{-4\taufill/t} /\Nbareq$.  This recovery of
$\phit(0)$ and the decay of $\vt$ imply that at time $\tauqs\approx
(\taufast \taufill)^{1/2}$ the two RHS terms in the velocity dynamics
eq. \eqref{milk} become equal.  Self-consistently, one finds that
thereafter they remain matched, \ie $\vtdot$ is much smaller than
either of these terms.  Thus from this time onwards, the fast velocity
variable (intrinsic timescale $\taufast$) evolves {\em
quasi-statically}, enslaved to the slow variable $\phit(0)$ (intrinsic
timescale $\tauslow$) according to
                                                \begin{eq}{vt}
\vt \approx -\vminus \phit(0) \gap 
				 (t>\tauqs) \period
                                                                \end{eq}
It follows that the velocity now undergoes a recovery as the hole
fills up. By $\taufill$ its magnitude is of order the equilibrium
value and can thus compete with diffusion. This heralds the onset of
episode (iii): we return to the basic dynamics, eq. \eqref{cc}, with
velocity and diffusion terms now of equal importance. For the first
time, relative perturbations of all quantities are now small, and we
can apply standard perturbation theory.  We find the $N=0$ chains
recover as $\delta\phit(0) \twid t^{-1/2}$ up to $\tauslow$ and $\twid
e^{-t/\tauslow}$ for $t>\tauslow$.  During these very late stages the
fast variables, which long ago decayed close to equilibrium, are
fine-tuned to their true equilibrium values, following $\phit(0)$
quasi-statically according to eq. \eqref{vt}.

Note we implicitly assumed sufficient time for the hole to develop
before mass transfer is complete, \ie $\taufast>\tstar$. This is true
provided $\epsilon>\epsilon_c=1/\Nbareq^{1/2}$.  For very small
perturbations $\epsilon<\epsilon_c$ the relative depth of the hole
though much greater than $\epsilon$ is less than unity.  Finally, we
have studied $\Delta T<0$. The response to a {\em positive} $T$-jump
is similar, but instead of a hole a large peak $\twid \Nbareq
\epsilon^2$ develops for small $N$, decaying after $\taufill$.

We tested our theory by numerical integration of coupled ODEs
describing the monomer-polymer dynamics. Fig. \ref{mwd} shows the
predicted deep hole developing after a small negative $T$-jump. Long
time enslavement of fast variables to slow ones is clearly
demonstrated (inset).

\begin{figure}
\includegraphics[width=7cm]{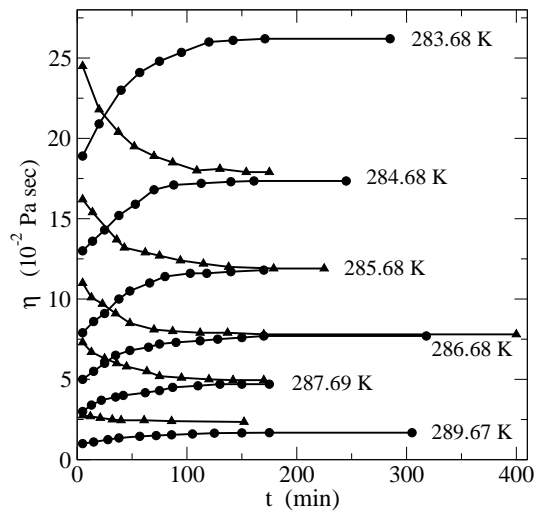}
\caption{\label{viscosity} 
Viscosity relaxation measurements by Ruiz-Garcia and Greer
\cite{garcia:greer:living_nonumber} following small $T$-jumps
($|\Delta T| \approx 1^0$K) on PAMS in tetrahydrofuran initiated by
sodium naphthalide.  Reproduced from fig. 1. of
ref. \onlinecite{garcia:greer:living_nonumber}.  Circles (triangles)
refer to positive (negative) $\Delta T$.  $\mtotal = 0.29$
gm/cm\cubed,  $\rhochains \approx 2.7 
\times 10^{-4}$ gm/cm\cubed (semi-dilute conditions). 
}
                                                   \end{figure}

We conclude by discussing our results and comparing to experiment. 
\ignore{ (This is a crucial point; systems with non-conserved numbers
of chains, including end-polymerizing wormlike surfactant micelles,
have very different dynamics as a result
\cite{livingpol_terminology_note}.)  }  
We found that, following a perturbation, living polymers relax by
adjusting (a) their total {\em mass}, proportional to first moment
$\Nbar$ and (b) the {\em shape} of their CLD. Mass variables are fast
(relaxation time $\taufast \twid \Nbar$) because monomer-polymer mass
transfer is a coherent process. Coherent chain growth, however, cannot
affect CLD shape whose relaxation therefore relies on slow
diffusion-like fluctuations in chain growth rates (relaxation time
$\tauslow \twid \Nbar^2$.) These timescales are typically separated by
2 or 3 orders of magnitude.  In fig. \ref{viscosity} we reproduce
viscosity relaxation measurements by Ruiz-Garcia and Greer
\cite{garcia:greer:living_nonumber} after small positive and negative
temperature jump perturbations of the living polymer PAMS at 6
different temperatures.  Now generally we expect viscosity $\eta \twid
c_\phi \Nbar^{\gamma}$ where $\gamma$ is a characteristic exponent
\cite{garcia:greer:living_nonumber} and $c_\phi$ depends on the entire
CLD, $\phi(N)$, including shape properties. Our theory thus predicts a
fast initial relaxation of $\eta$ in a time $\taufast$ followed by a
very slow relaxation in $\tauslow$.  Now
\cite{garcia:greer:living_nonumber} $\rhochains \approx 2.7 \times
10^{-4}$ gm/cm\cubed \ while \cite{zhuang:greer_kinetics_etal} $\kplus
\approx 0.2 M^{-1}$ sec$^{-1}$ was measured to vary by $\approx 10\%$
over this temperature range. Thus, rewriting
$\taufast=1/\rhochains\kplus$, we predict a {\em constant} relaxation
time ($\pm 10\%$) for all 12 measurements, $\taufast\approx 2000$
seconds.  This agrees very closely with experiment (see
fig. \ref{viscosity}) despite the fact that the observed viscosities
varied by an order of magnitude and the mean chain lengths are
estimated to vary from almost zero up to several thousand units at the
lowest temperature.  Note $\tauslow = \taufast^2 \vminus\theta \approx
$ 1 week-1 month, after estimating \cite{greer:review} $\theta\approx
0.5$ and using $\vminus=0.1$ sec$^{-1}$. The second shape-derived
relaxation of $\eta$ is therefore unobservably long for this
experiment.

Very large temperature quenches were studied in
ref. \onlinecite{das:greer:living_7_mwd_etal}. $\Nbar$ relaxed faster
than the second moment, though it was unclear if $m$ and $\Nbar$ were
coupled as required by mass conservation, possibly due to ionic
aggregation\cite{frischknechtmilner:living_aggregation}. The authors
of ref. \onlinecite{miyakestockmayer:living} studied large
perturbations theoretically, identifying the timescales $\taufast$ and
$\tauslow$. Their analytical results were valid for short times and
very small $\theta$.

In this letter we showed that even a small perturbation leads to a
non-linear dynamical response.  This ultrasensitivity is due to the
inability of slow CLD shape variables to keep pace with the fast
relaxation of $\Nbar$ which entails simple translation of the CLD
leaving a hole or peak at small N.  Of all slow variables, the most
sensitive is $\phit(0)$, the number of $N=0$ chains, which becomes
exponentially small or massively enhanced. These free initiators also
govern the very late fine-tuning of the fast variables $\Nbar$ and
$m$, the free monomer concentration.  Physically, this is because
monomers add to chain ends only.  But all ends are identical except
for the $N=0$ chains which cannot shed monomers.  Thus $\phit(0)$ is
the only dynamic polymer property featuring in the kinetics of
$m$. Given its central role, we propose measurement of the number of
free initiators by spectroscopic or other methods as a revealing probe
of ultrasensitivity.


This work was supported by the Petroleum Research Fund under grant
no. 33944-AC7. 




\begin{thebibliography}{10}
\bibitem{living:kplusvminus_independentofN_note}
We assume $\kplus, \vminus$ independent of N, consistent with the ideal chain
  statistics of semidilute solutions or
  melts\cite{schafer:living_theory_prl_combo}.
\end{thebibliography}

\end{document}